\documentclass[aps,pra,showpacs,twocolumn,preprintnumbers,superscriptaddress]{revtex4}
\usepackage{amsmath}
\usepackage{amssymb}
\usepackage{graphicx}

\usepackage{color}
\usepackage{bm}
\usepackage{epstopdf}

\begin{document}

\title{Asymmetric quantum dot in microcavity as a nonlinear optical element}

\author{I. G. Savenko}
\affiliation{Science Institute, University of Iceland, Dunhagi 3,
IS-107, Reykjavik, Iceland}

\author{O. V. Kibis}
\affiliation{Department of Applied and Theoretical Physics,
Novosibirsk State Technical University, Karl Marx Avenue 20,
630092 Novosibirsk, Russia}

\author{I. A. Shelykh}
\affiliation{Science Institute, University of Iceland, Dunhagi 3,
IS-107, Reykjavik, Iceland} \affiliation{Division of Physics and
Applied Physics, Nanyang Technological University 637371,
Singapore}

\begin{abstract}
We have investigated theoretically the interaction between
individual quantum dot with broken inversion symmetry and
electromagnetic field of a single-mode quantum microcavity. It is
shown that in the strong coupling regime the system demonstrates
nonlinear optical properties and can serve as emitter of the
terahertz radiation at Rabi frequency of the system. Analytical
results for simplest physical situations are obtained and
numerical quantum approach for calculating emission spectrum is
developed.
\end{abstract}


\maketitle


\section{Introduction}

Quantum microcavity is unique laboratory for studies of the strong
light-matter coupling. Being first observed two decades ago
\cite{Weisbuch}, the strong coupling regime is now routinely
achieved in different kinds of microcavities \cite{Microcavities}.
From the fundamental viewpoint, it is interesting as a basis to
investigate various collective phenomena in condensed matter
systems such as Bose-Einstein condensation (BEC) \cite{BEC} and
superfluidity \cite{Superfluidity}. From the viewpoint of
applications, it opens a way to the realization of optoelectonic
devices of the new generation \cite{OptApplications}:
room-temperature polariton lasers \cite{PolLasers}, polarization-
controlled optical gates \cite{OptGates}, and others.

Several applications of the strong coupling regime were also
proposed for the quantum information processing
\cite{Savasta,QInformatics,Johne}. In this case one should be able
to tune the number of emitted photons in controllable way. This is
hard to achieve in planar microcavities where the number of
elementary excitations is macroscopically large, but is possible
in microcavities containing single quantum dots (QDs) where QD
exciton can be coupled to a confined electromagnetic mode provided
by a micropillar (etched planar cavity)
\cite{QDStrongCouplingPillar}, a defect of the photonic crystal
\cite{QDStrongCouplingDefect} or a whispering gallery mode
\cite{QDStrongCouplingWGM}. That is why the strong coupled systems
based on QDs have attracted particular attention recently.

Depending on the size of the QD, its elementary excitations can
behave as fermions (small QDs whose size is comparable with
exciton radius in the bulk material) \cite{LibLaussyFermions},
bosons (large QDs whose size is much larger than exciton radius in
the bulk material) \cite{LibLaussyBosons} or particles with
intermediate statistics (medium size QDs) \cite{LaussyMultiplets}.
In the current paper, we consider a small single QD in a
microcavity that corresponds to the case of fermions. For
symmetric QDs such system can be described by the well-known
Jaynes-Cummings Hamiltonian \cite{Jaynes_63} which predicts
transformation of the Rabi doublet to the Mollow triplet in the
emission spectrum
 as intensity of the external pump growth for both coherent \cite{Mollow} and incoherent excitation schemes \cite{LaussyMollow}. On the other hand,
the incorporation of the asymmetry into the quantum system can
radically change its emission pattern and lead to the opening of
optical transitions which were forbidden in the symmetric case.
Particularly, the breaking of inversion symmetry opens optical
transitions at the Rabi frequency
 at QDs placed in strong external laser field \cite{LibKibis}.
The similar effect occurs for asymmetric quantum wells placed
inside a planar microcavity \cite{LibShelykhTHz, LibSavenkoTHz}.
In current manuscript we consider modification of the emission
spectrum of asymmetric QDs inside a single-mode microcavity using
fully quantum approach.

The work is organized as follows. In Section II we describe the
formalism and introduce the model Hamiltonian. In Section III we
obtain analytical solutions for the important particular cases. In
Section IV we discuss the incorporation of pump and decay terms
into the Hamiltonian and present the numerical calculations of the
emission spectrum. Section V contains discussions and conclusions.


\section{Model}

We model QD as a two-level quantum system with the ground state
$|g\rangle$ and excited state $|e\rangle$ with energies
$\varepsilon_g$ and $\varepsilon_e$, respectively. QD is placed
inside a cavity and interacts resonantly with confined
electromagnetic mode of the frequency $\omega_c$.  Since
electromagnetic field can transfer an electron in the QD from the
valence band into the conduction band, the ground state
$|g\rangle$ corresponds to the absence of free carriers while the
first excited state $|e\rangle$ is the state with an electron in
the conduction band and a hole in the valence band. Therefore, the
energy difference $\Delta=\varepsilon_g-\varepsilon_e$ is
approximately equal to the band gap of the QD minus excitonic
correction accounting for the Coulomb attraction between the
electron and the hole.

The full Hamiltonian of the system can be represented as a sum of
three parts,
\begin{equation}
\label{H} {\hat{{\cal H}}}=\hat{{{\cal
H}}}_{\mathrm{e}}+\hat{{{\cal H}}}_{\mathrm{\omega}}+\hat{{{\cal
H}}}_{\mathrm{int}}\,,
\end{equation}
where
\begin{equation}
\hat{{{\cal H}}}_{\mathrm{e}}=\frac{\Delta}{2}\sigma_z
\end{equation}
is the Hamiltonian of the single QD, and $\sigma_{x,y,z}$ are the
Pauli matrices acting in the space of $|e\rangle$ and $|g\rangle$
states. The Hamiltonian of the free electromagnetic field reads
\begin{equation}
\hat{{{\cal H}}}_{\mathrm{\omega}}=\hbar\omega_c{a}^\dagger{a},
\end{equation}
where $a,a^\dagger$ are the annihilation and creation operators
for cavity photons, respectively. The Hamiltonian $\hat{{\cal
H}}_{\mathrm{int}}$ describes interaction of the QD with the
electromagnetic field and can be constructed as following. The
interaction of a classical dipole $\mathbf{d}$ with a classical
external electric field $\mathbf{E}$ is given by the expression
$\hat{{{\cal H}}}_{\mathrm{int}}=-\mathbf{E}\mathbf{d}$.
\begin{figure}[th]
\includegraphics[width=0.48 \textwidth]{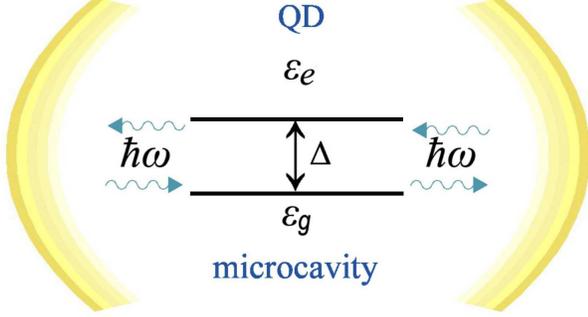}
\caption{Two-level quantum dot with the bandgap $\Delta$ in a
single-mode microcavity with the frequency $\omega_c$.}
\label{FigSystem}
\end{figure}
Within the quantum-field approach, we have to replace the
classical quantities $\mathbf{d}$ and $\mathbf{E}$ with the
corresponding operators
\begin{eqnarray}
\hat{\textbf{E}}&=&\sqrt{\frac{\hbar\omega_c}{2\epsilon_0
V}}(\mathbf{e}a+\mathbf{e}^\ast a^\dagger),\\
\nonumber \hat{\textbf{d}}&=&\left(\begin{array}{cc}
  \textbf{d}_{ee} & \textbf{d}_{eg}\\
  \textbf{d}_{ge} & \textbf{d}_{gg}\\
\end{array}\right)=\frac{\textbf{d}_{ee}+\textbf{d}_{gg}}{2}I+
\frac{\textbf{d}_{ee}-\textbf{d}_{gg}}{2}\sigma_z\\
\nonumber
&&~~~~~~~~~~~~~~~~~~+(\textbf{d}_{eg}\sigma^++\textbf{d}_{ge}\sigma^-),
\end{eqnarray}
where $\mathbf{e}$ is the polarization vector of the cavity mode,
$\epsilon_0$ is the vacuum dielectric permittivity,
$\textbf{d}_{ij}=\langle i|\hat{\textbf{d}}|j\rangle$ are dipole
matrix elements of the QD, $I$ is the unity matrix,
$\sigma^\pm=(\sigma_x\pm i\sigma_y)/2$. In the case of symmetric
QD, the matrix elements $\textbf{d}_{ee}$ and $\textbf{d}_{gg}$
are zero. Due to the breaking of inversion symmetry in asymmetric
QDs, the dipole matrix elements appear to be nonequivalent,
$\textbf{d}_{ee}\neq\textbf{d}_{gg}$. This leads to new physical
effects discussed hereafter. The interaction Hamiltonian can be
written as
\begin{eqnarray}
\nonumber
\hat{{\cal H}}_{\mathrm{int}}&=&-\hat{\textbf{d}}\hat{\textbf{E}}\\
\nonumber
&=&g_R\left(a+a^\dagger\right)(\sigma^++\sigma^-)+g_S\left(a+a^\dagger\right)\left({\sigma}_z+I\right)\\
&\approx& g_R\left(a\sigma^++a^\dagger\sigma^-\right)+
g_S\left(a+a^\dagger\right)\left({\sigma}_z+I\right).
\label{EqHamiltonian}
\end{eqnarray}
The coupling parameters,
$g_R=-(\textbf{d}_{eg}\cdot\textbf{e})\sqrt{{\hbar}\omega_c/{2\epsilon_0
V}}$ and $g_S=-(\textbf{d}_{ee}\cdot
\textbf{e})\sqrt{{\hbar}\omega_c/4\epsilon_0{V}}$, describe the
excitation-exchange interaction and the symmetry-dependent
interaction, respectively. For definiteness, we assume them to be
real. The parameter $V$ in the expressions above is the
quantization volume and can be estimated as
$V\approx(\lambda/2)^3$, where $\lambda=c/2\pi\omega_c$ is the
characteristic wavelength corresponding to the cavity mode. We
also put $\textbf{d}_{gg}=0$, which can be justified for
nonferroelectric QDs. Indeed, asymmetry of such QDs is provided by
peculiar shape and/or an external electric field. Due to it, the
matrix element $d_{gg}$ is proportional to the size of the
elementary cell of the crystal lattice, while the matrix element
$d_{ee}$ is proportional to the size of the QD. As a result, in
realistic QDs one has $d_{ee}\gg d_{gg}$. To pass from the second
line to the third line in Eq.~(\ref{EqHamiltonian}), the
rotating-wave approximation \cite{Scully_b01,Cohen-Tannoudji_b98}
was applied and we dropped the anti- resonant terms proportional
to $a^\dagger\sigma^+$ and $a\sigma^ -$.

The full Hamiltonian of the system reads
\begin{eqnarray}
\nonumber
\hat{{\cal H}}=\hbar\omega_c{a}^\dagger{a}+\frac{\Delta}{2}\sigma_z\\
+g_R\left(a\sigma^++a^\dagger\sigma^-\right)+
g_S\left(a+a^\dagger\right)\left({\sigma}_z+I\right).
\label{EqFullHamiltonian}
\end{eqnarray}
In the case of a symmetric QD, the coupling parameter $g_S$ equals
zero. Then Eq.~(\ref{EqFullHamiltonian}) reduces to the
Hamiltonian of the fully solvable Jaynes-Cummings model. Its
eigenstates correspond to electronic excitations dressed by the
cavity photons and can be expressed as
\begin{equation}
|\psi_n^{\pm(0)}\rangle=A_n^{\pm}|g,n\rangle+B_n^{\pm}|e,n-1\rangle
\label{EqWavefunction0},
\end{equation}
where
\begin{equation}
A_n^{\pm}=\frac{\varepsilon_n^{\pm(0)}-\hbar\omega_c(n-1)-\Delta/2}
{\sqrt{[\varepsilon_n^{\pm(0)}-\hbar\omega_c(n-1)-\Delta/2]^2+g_R^2n}},
\label{An}
\end{equation}
\begin{equation}
B_n^{\pm}=\frac{g_R\sqrt{n}}{\sqrt{[\varepsilon_n^{\pm(0)}-\hbar\omega_c(n-1)-\Delta/2]^2+g_R^2n}},
\label{Bn}
\end{equation}
and the composite electron-photon states
$|g,n\rangle=|g\rangle\bigotimes|n\rangle$ and
$|e,n\rangle=|e\rangle\bigotimes|n\rangle$ describe both the QD
state (the ground state $g$ or the excited state $e$) and the
field state with $n$ cavity photons. It should be noted that the
Jaynes-Cummings Hamiltonian commutes with the excitation number
operator
\begin{equation}
\hat{N}=a^\dagger a+(\sigma_z+I)/2, \label{EqExNumOp}
\end{equation}
whose eigenstates correspond to the conserved number of total
electron-photon excitations in the system counted as number of the
excitations in QD (one for the state $|g\rangle$, zero for the
state $|e\rangle$) plus number of the photons in the cavity mode.
The eigenenergies corresponding to the states
(\ref{EqWavefunction0}) are given by
\begin{equation}
\varepsilon_n^{\pm(0)}=\hbar\omega_c\left(n-\frac{1}{2}\right)\pm\sqrt{\frac{(\hbar\omega_c-\Delta)^2}{4}+g_R^2n}.
\label{EqEigenEn}
\end{equation}
In order to obtain the emission spectrum of the system, we need to
analyze optical transitions between the eigenstates
(\ref{EqWavefunction0}). An emitted photon goes outside the
system, and a pumped photon appears in the system. Thus, there is
an exchange of photons between the coupled QD-microcavity system
and some external reservoir. Therefore, we can introduce the
Hamiltonian of the exchange of photons between cavity and outside
world,
\begin{equation}
\hat{\cal H}_{ex}=\hbar(\Gamma^\star a r^\dagger+\Gamma a^\dagger
r), \label{EqHex}
\end{equation}
where $r^\dagger,r$ are creation and annihilation operators for
the external photons, $\Gamma$ is the system-reservoir coupling
constant.

The probabilities (intensities) of transitions with emission of
photon from the cavity are proportional to the corresponding
matrix elements,
\begin{equation}
I_{if}\sim|\langle\psi_f,1_R|\hat{\cal
H}_{ex}|\psi_i,0_R\rangle|^2, \label{Iif}
\end{equation}
where the symbols $\psi_f,\psi_i$ denote the final and initial
eigenstates of the Hamiltonian (\ref{EqWavefunction0}), $0_R$ and
$1_R$ describe zero- and one-photon states of the reservoir.
Substituting Eq.~(\ref{EqHex}) into Eq.~(\ref{Iif}), one gets
\begin{eqnarray}
\label{EqTrans} \nonumber
I_{if}\sim|\langle\psi_f|a|\psi_i\rangle|^2=\\
\label{EqTransUnp}
=\left(\sqrt{n_i}A^\pm_{n_f}A^\pm_{n_i}+\sqrt{n_f}B^\pm_{n_f}B^\pm_{n_i}\right)^2\delta_{n_f,n_i-1},\label{Iif1}
\end{eqnarray}
where the states $|\psi_{i,f}\rangle$ are defined by
Eq.~(\ref{EqWavefunction0}), and integers $n_i,n_f$ are initial
and final numbers of electron-photon excitations in the system,
defined earlier (see Eq.~(\ref{EqExNumOp})).

The Kronecker delta in Eq.~(\ref{EqTrans}) means that only
transitions changing number of excitations by one are allowed in a
system described by the Jaynes-Cummings Hamiltonian. It follows
from Eq.~(\ref{EqEigenEn}) that in the case of resonance
($\hbar\omega_c=\Delta$) the optical spectrum contains peaks at
the energies $\hbar\omega_c\pm g_R(\sqrt{n+1}\pm\sqrt{n})$, where
$n=0,1,2,...$. The accounting of the peaks broadening leads to the
emission spectrum in the form of the Rabi doublet (for $n=0$), and
the Mollow triplet (for $n\gg1$) \cite{LaussyMollow}. In the
intermediate regime more complicated multiplet structure can be
observed \cite{LibLaussyFermions}.

It should be noted that in symmetric QDs ($g_S=0$) transitions at
the Rabi frequency $\Omega_R=2g_R/\hbar$ are forbidden. Indeed,
the transitions would occur between electron-photon states with
the same number of excitations, $n_f=n_i$, that is not allowed by
Eq.~(\ref{Iif1}). However, these transitions become possible for
asymmetric QDs, i.e. when the full Hamiltonian
(\ref{EqFullHamiltonian}) contains the term
$g_S\left(a+a^\dagger\right)\left({\sigma}_z+1\right)$
\cite{LibKibis}. Since for realistic microcavities the Rabi
frequency $\Omega_R$ lies in the terahertz (THz) range, such
transitions form the physical basis for using the considered
system as a tunable source of THz radiation. We will consider
these THz transitions in more details in the following sections.
%

\section{Analytical solutions}

First of all, let us consider analytically the case of weak
asymmetry. Then the Hamiltonian (\ref{EqFullHamiltonian}) can be
represented as a sum of two parts
\begin{eqnarray}
\hat{\cal{H}}=\hat{\cal{H}}_{JC}+\widehat{V},
\end{eqnarray}
where
\begin{equation}
\hat{\cal{H}}_{JC}=\hbar\omega_c a^\dagger
a+\frac{\Delta}{2}{\sigma}_z+g_R(a\sigma^+ +a^\dagger\sigma^-)
\label{EqJC}
\end{equation}
is the Jaynes-Cummings Hamiltonian, and
\begin{equation}
\widehat{V}={g}_S(a+a^\dagger)({\sigma}_z+I) \label{EqV}
\end{equation}
is the term arising from the asymmetry of the QD. Considering the
term (\ref{EqV}) as a small perturbation and using the standard
first-order perturbation theory, the corrected wavefunctions of
the system can be written as
\begin{equation}
\label{EqWfPerturb}
|\psi_n^{\pm}\rangle=|\psi_n^{\pm(0)}\rangle+\sum_{m}\sum_{\alpha=\pm}{\frac{V_{nm}^{\pm\alpha}}{\varepsilon_n^{\pm(0)}-\varepsilon_m^{\alpha(0)}}}|\psi_m^{\alpha(0)}\rangle,
\end{equation}
where $|\psi_n^{\pm(0)}\rangle$ are the unperturbed eigenfunctions
(\ref{EqWavefunction0}), $\varepsilon_n^{\pm(0)}$ are the
corresponding eigenenergies (\ref{EqEigenEn}), and
$V_{nm}^{\pm\alpha}=\langle\psi_n^{\pm(0)}|\widehat{V}|\psi_m^{\alpha(0)}\rangle$
are the matrix elements of the perturbation (\ref{EqV}).

It is easy to see that these matrix elements are different from
zero only if $m=n\pm1$, and thus states with $n,n-1$ and $n+1$
excitations become mixed by the asymmetry of QD. This mixture
leads to the opening of the optical transitions
$|\psi_n^{+}\rangle\rightarrow|\psi_n^{-}\rangle$ at the
frequencies $\Omega_R\sqrt{n}$. As well, transitions at double
frequencies of the cavity $2\omega_c\pm
\Omega_R(\sqrt{n+1}\pm\sqrt{n})/2$ become opened. The allowed
transitions are shown schematically in Fig.~(2).
\begin{figure}[th]
\includegraphics[width=0.48 \textwidth]{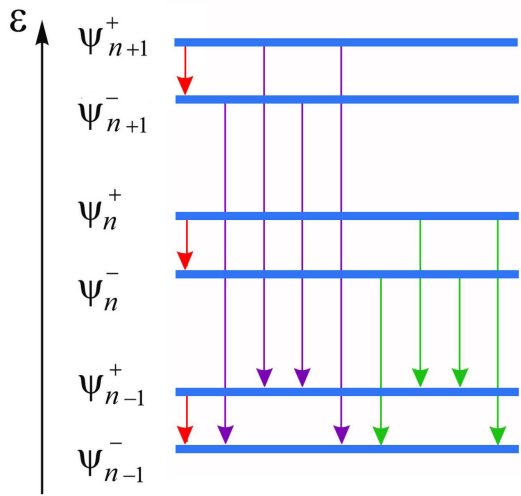}
\caption{Optical transitions in the asymmetrical QD coupled with
the cavity mode.} \label{FigStates}
\end{figure}

Since Rabi frequency $\Omega_R$ lies typically in the THz range,
asymmetric QD-cavity system can be used as a nonlinear THz
emitter. Using Eqs.~(\ref{EqTrans}) and (\ref{EqWfPerturb}), we
obtain the intensity of THz transitions as follows:
\begin{equation}
\nonumber I_{if}\sim|\langle \psi_n^-|a|\psi_n^+\rangle|^2,
\label{IR}
\end{equation}
where
\begin{eqnarray}
\nonumber
\langle \psi_n^-|a|\psi_n^+\rangle=~~~~~~~~~~~~~~~~~\\
\nonumber
=\frac{V_{n-1,n}^{-+}}{\varepsilon_{n}^{-(0)}-\varepsilon_{n-1}^{+(0)}}\left(\sqrt{n}A_{n-1}^+A_n^++\sqrt{n-1}B_{n-1}^+B_n^+\right)\\
\nonumber
+\frac{V_{n+1,n}^{+-}}{\varepsilon_{n}^{+(0)}-\varepsilon_{n+1}^{-(0)}}\left(\sqrt{n+1}A_{n}^-A_{n+1}^-+\sqrt{n}B_{n}^-B_{n+1}^-\right).
\end{eqnarray}
with coefficients $A_n^ \pm$ and $B_n^\pm$ given by expressions
(\ref{An}) and (\ref{Bn}), respectively.



One can also consider analytically another physical situation: the
asymmetry is no longer assumed to be a weak perturbation, but
photon occupation numbers are supposed to be large, $n\gg1$ . This
limiting case corresponds to the classical electromagnetic field
in the cavity. Let us represent the full Hamiltonian
(\ref{EqFullHamiltonian}) as a sum of the ``diagonal'' part
\begin{equation}
\label{EqHd} \hat{{\cal H}}_d=\hbar\omega_c a^\dagger a+
\frac{\Delta}{2}{\sigma}_z+g_S\left(a+a^\dagger\right)\left({\sigma}_z+I\right)
\end{equation}
and the ``off-diagonal'' part
\begin{equation}
\label{EqHnd} \hat{{\cal
H}}_{od}=g_R\left(a\sigma^++a^\dagger\sigma^-\right).
\end{equation}
The Hamiltonian (\ref{EqHd}) does not commute with the excitation
number operator (\ref{EqExNumOp}), but commutes with the Pauli
matrix $\sigma_z$. This means that eigenstates of the Hamiltonian
(\ref{EqHd}) can be represented as
\begin{eqnarray}
|\psi_n^{g}\rangle=\sum_{k=0}^{\infty}C^{g}_{nk}|g,k\rangle,\label{psi1}\\
|\psi_n^{e}\rangle=\sum_{k=0}^{\infty}C^{e}_{nk}|e,k\rangle.\label{psi2}
\end{eqnarray}
After substituting the expressions (\ref{psi1}), (\ref{psi2}) into
the Schr\"odinger equation $\hat{{\cal
H}}_d|\psi^{g,e}_{n}\rangle=\varepsilon^{g,e}_{n}|\psi^{g,e}_{n}\rangle$
with the Hamiltonian (\ref{EqHd}), we obtain the system of
algebraic equations for coefficients $C^{g,e}_{nk}$:
\begin{equation}
\label{EqCn1} \left[\hbar\omega_c
k-\Delta/2-\varepsilon^g_n\right] C_{nk}^{g}=0
\end{equation}
for $k=0,1,2,..$,
\begin{equation}
\left[\hbar\omega_c k+\Delta/2-\varepsilon^e_n\right] C_{nk}^{e}
+2g_S\sqrt{k+1}C_{n,k+1}^{e}=0\label{EqCn2}
\end{equation}
for $k=0,1$,
\begin{eqnarray}
\left[\hbar\omega_c k+\Delta/2-\varepsilon^e_n\right]
C_{nk}^{e}\nonumber\\
+2g_S\left(\sqrt{k+1}C_{n,k+1}^{e}+\sqrt{k}C_{n,k-1}^{e}\right)=0\label{EqCn3}
\end{eqnarray}
for $k=2,3,4,..$.

The solutions of Eqs.~(\ref{EqCn1}) are evident:
\begin{eqnarray}
\nonumber
\varepsilon_n^g=\hbar\omega_c n-\frac{\Delta}{2},\\
C^{g}_{nk}=\delta_{nk}.
\end{eqnarray}
As for  Eqs.~(\ref{EqCn3}), in the limiting case $k\gg1$ they are
similar to the well-known recurrent expression for the Bessel
function of the first kind:
\begin{equation}
\label{Z} 2mJ_m(x)=xJ_{m-1}(x)+xJ_{m+1}(x),
\end{equation}
where $m$ is an integer and $x$ is the argument of the Bessel
function of the first kind, $J_m(x)$. Therefore, we can write the
solutions of Eqs.~(\ref{EqCn3}) for $k\gg1$ as

\begin{eqnarray}
\nonumber
\varepsilon_n^e=\hbar\omega_c n+\frac{\Delta}{2},\\
C^{e}_{nk}=J_{k-n}(x_k)\,,\label{S}
\end{eqnarray}
where $x_k=-4g_S\sqrt{k}/\hbar\omega_c$. It should be noted that
the solutions (\ref{S}) satisfy Eqs.~(\ref{EqCn2})--(\ref{EqCn3})
for small integers $k\sim1$ as well, since
$$\lim_{n\rightarrow\pm\infty}J_n(x)=0.$$
As a result, for large photon occupation numbers $n$ the
eigenfunctions (\ref{psi1})--(\ref{psi2}) take the form
\begin{eqnarray}
|\psi_n^{g}\rangle&=&|g,n\rangle,\label{psi11}\\
|\psi_n^{e}\rangle&=&\sum_{k=0}^{\infty}J_{k-n}(x_k)|e,k\rangle.\label{psi22}
\end{eqnarray}

Let us make $n$ and $V$ tend to infinity while keeping $n/V$
constant. This limiting case corresponds to the conventional model
of an intense laser-generated field \cite{CT}. Then
$x_n=(\textbf{d}_{ee}\cdot \textbf{e})E_n/\hbar\omega_c$, where
$E_n=\sqrt{2n\hbar\omega_c/\epsilon_0 V}$ is the classical
amplitude of the field. Therefore, in the case of the most
relevant physical situation with $x_n\ll1$, the eigenfunctions
(\ref{psi22}) can be estimated as
$|\psi_n^{e}\rangle\approx|e,n\rangle$ since
$J_{k-n}(0)=\delta_{k,n}$. This allows to seek eigenfunctions of
the full Hamiltonian $\hat{{\cal H}}=\hat{{\cal H}}_{d}+\hat{{\cal
H}}_{od}$ as a superposition (linear combination) of the functions
(\ref{psi11}) and (\ref{psi22}). Substituting this superposition
$\psi^\pm_{n}$ into the Schr\"odinger equation $\hat{\cal
H}|{\psi}^\pm_{n}\rangle={\varepsilon}^\pm_{n}|{\psi}^\pm_{n}\rangle$,
we can find the energy spectrum ${\varepsilon}^\pm_{n}$ of the
coupled electron-photon system. Particularly, for the resonance
case ($\Delta=\hbar\omega_c$) this linear combination is
\begin{equation}
\label{EqApPsi}
|{\psi}^\pm_{n}\rangle=\frac{1}{\sqrt{2}}(|\psi^{e}_{n}\rangle\pm|\psi^{g}_{n+1}\rangle),
\end{equation}
and the corresponding energies are
\begin{equation}\label{EqApEnergu}
\varepsilon^\pm_n=\hbar\omega_c
n+\varepsilon_e\pm\frac{\hbar\Omega_R^\prime}{2},
\end{equation}
where $\Omega_R^\prime=(\textbf{d}_{eg}\cdot \textbf{e})E_n/\hbar$
is the Rabi frequency for the classically strong electromagnetic
field. It follows from the expressions
(\ref{psi11})--(\ref{EqApPsi}) that the states with all possible
numbers of excitations become intermixed. This means that all
transitions $|\psi_n^{\pm}\rangle\rightarrow|\psi_m^{\pm}\rangle$
are allowed and emission spectrum contains frequencies
$(n-m)\hbar\omega_c\pm\Omega_R^\prime$. However, the intensity of
the transitions decreases with increasing $(n-m)$ because of
decreasing the Bessel functions $J_k(x)$ with increasing $k$
\cite{LibKibis}. Therefore, most intensive ones correspond to
those depicted at Fig.~\ref{FigStates}. This agrees with the
results obtained above in the frameworks of perturbation theory
for the weak asymmetry case.

It should be noted that there is no analytical solution for
arbitrary photon occupation numbers of the cavity mode and
arbitrary asymmetry strength. To study physically relevant
situations outside the obtained analytical solutions as well as to
calculate the shape of the emission spectrum, we need to apply the
numerical approach discussed in the next section.


\section{Numerical approach}

The discussion in the previous section was dedicated to the
analytical calculation of the energy spectrum of the system. In
two limiting cases corresponding to weak asymmetry and large
photon occupation numbers we were able to find the emission
frequencies. However, even in these cases our treatment did not
allow the calculation of the shape of the emission spectrum as
function of the intensity of the external pump. In this section we
calculate it numerically, using the approach based on the master
equation techniques.

Let us assume that one has a QD embedded in a microcavity and
switches on incoherent pumping of the photonic mode. After some
time an equilibrium is established and steady state (SS) is
reached. It means that the increase of photon number provided by
external pumping is balanced by escape of photons from the cavity.
In this regime, one can measure the emission spectrum, i.e. the
intensity of the flux of the photons going out from the cavity as
a function of their frequency.

In quantum optics, the standard way of consideration of the
processes involving external pumping and decay is based on using
of the master equation for the full density matrix of the system
$\rho$ (see, e.g., Ref.\cite{LibMicrocavities}) which can be
represented in the following form:
\begin{equation}
\partial_t\rho=\frac{1}{i\hbar}[{\hat{\cal H}};\rho]+{\cal L}\rho.
\label{EqMaster}
\end{equation}
The first term on the right-hand side of the equation, $\hat{\cal
H}$, stands for the Hamiltonian describing coherent processes in
the system, and the symbol $\cal L$ denotes the Lindblad
superoperator accounting for pump and decay. In the case we
consider, $\hat{\cal H}$ is given by the expression
(\ref{EqFullHamiltonian}), while the Lindblad term reads:
\begin{eqnarray}
\nonumber
{\cal L}\rho&=&P\left(a\rho a^\dagger+a^\dagger\rho a-a^\dagger a\rho-\rho aa^\dagger\right)\\
\nonumber &+&\frac{\gamma_{ph}}{2}\left(2a\rho
a^\dagger-\rho a^\dagger a-a^\dagger a\rho\right)\\
&+&\frac{\gamma_{QD}}{2}\left(2\sigma\rho \sigma^+-\rho \sigma^+
\sigma-\sigma^+ \sigma\rho\right).
\end{eqnarray}
Here $P$ is the intensity of the incoherent pump of the cavity
mode, $\gamma_{ph}$ and $\gamma_{QD}$ are broadenings of photonic
and excitonic modes, respectively (the latter is taken zero in all
calculations below). Equation ~(\ref{EqMaster}) represents a set
of linear ordinary differential equations for matrix elements of
the density matrix $\rho$. In our numerical analysis we will use
the basis
\begin{equation}
|g,n\rangle,  ~~~ |e,n\rangle \label{EqMasterBasis}
\end{equation}
that gives us a following system of equations, which can be
written briefly as
\begin{equation}
\label{EqLindbladMatrix}
\partial_t\rho_{ij}^{ab}=M_{ijkl}^{abcd}\rho_{kl}^{cd},
\end{equation}
or, in the explicit form, as
\begin{widetext}
\begin{eqnarray}
\label{EqLindbladMatrixExplicit}
\partial_t\rho_{ij}^{ab}&=&\hbar\omega_c(i-j)\rho_{ij}^{ab}+\frac{\Delta}{2} \left[(-1)^{\delta_{ag}}-(-1)^{\delta_{bg}}\right]\rho_{ij}^{ab}\\
\nonumber &+&g_R\left(\sqrt{i+1}\delta_{ae}\rho_{i+1,j}^{ab}
              +\sqrt{i}     \delta_{ag}\rho_{i-1,j}^{ab}
              - \sqrt{j}\delta_{bg}\rho_{i,j-1}^{ab}
              - \sqrt{j+1}\delta_{be}\rho_{i,j+1}^{ab}\right)\\
\nonumber &+&g_S\left(\sqrt{i+1}\delta_{ae}\rho_{i+1,j}^{ab}
              +\sqrt{i}     \delta_{ae}\rho_{i-1,j}^{ab}
              - \sqrt{j}\delta_{be}\rho_{i,j-1}^{ab}
              - \sqrt{j+1}\delta_{be}\rho_{i,j+1}^{ab}\right)\cdot 2\\
\nonumber &+&\frac{P}{2}
                         \left(2\sqrt{(i+1)(j+1)}\rho_{i+1,j+1}^{ab}
                               +2\sqrt{ij}\rho_{i-1,j-1}^{ab}
                               - (2i+2j+2)\rho_{i,j}^{ab}\right)\\
\nonumber &+&\frac{\gamma_{ph}}{2}
                              \left(2\sqrt{(i+1)(j+1)}\rho_{i+1,j+1}^{ab}
                                   - (i+j)\rho_{i,j}^{ab}\right)
     +\frac{\gamma_{QD}}{2}
                              \left(2\rho_{i,j}^{ab}\delta_{ag}\delta_{bg}
                  -\rho_{i,j}^{ab}\delta_{ae}-\rho_{i,j}^{ab}\delta_{be}\right).
\nonumber
\end{eqnarray}
\end{widetext}
Here the superscripts $a,b,c,d$ are either $g$ or $e$ (ground or
excited state of the QD) and subscripts $i,j,k,l$ correspond to
the number of the photons in a cavity. In principle, they can take
values from zero to infinity, but for numerical analysis
truncation of the matrix is needed. The stronger is the pump, the
more states should be taken into account. The natural way to
control the accuracy of the truncation is to check the
conservation of the trace of the truncated density matrix.

Numerical solution of the system (\ref{EqLindbladMatrix}) allows
to find the full density matrix of the system in stationary state,
$\rho_{ij}^{SS}$, which allows to determine the probabilities of
the occupancies of different quantum states of the coupled QD-
cavity system as functions of the intensity of the external pump.
As well, it allows to find the shape of the emission spectrum of
the system. To pursue this latter task, we will use the two
approaches. Let us start with the relatively simple model based on
the modified Fermi golden rule, considering the isolated QD-
cavity system. Its eigenstates can be found by diagonalization of
the Hamiltonian (\ref{EqFullHamiltonian}). This procedure was
performed analytically in the previous section for the cases of
weak asymmetry and large photonic occupation numbers. However, in
general case a numerical analysis is needed. Let the system is in
a pure state corresponding to one of its eigenstates. Then,
according to the Fermi golden rule, we can estimate the emission
spectrum as
\begin{eqnarray}
&&S(\omega)\sim\label{Iomega1}\\
\nonumber
&\sim&\sum_f|\langle\psi_f,1_R|{\hat{\cal H}}_{ex}|\psi_i,0_R\rangle|^2\wp(\omega)\frac{\gamma_{ph}^2}{(\varepsilon_f-\varepsilon_i-\hbar\omega)^2+\gamma_{ph}^2}\\
\nonumber
&\sim&|a_{if}|^2\wp(\omega)\frac{\gamma_{ph}^2}{(\varepsilon_f-\varepsilon_i-\hbar\omega)^2+\gamma_{ph}^2},\nonumber
\end{eqnarray}
where the symbols $\psi_f,\psi_i$ denote the final and initial
eigenstates of the Hamiltonian (\ref{EqFullHamiltonian}), $0_R$
and $1_R$ describe zero- and one-photon states of the reservoir,
$\hat{\cal H}_{ex}$ is the Hamiltonian of the coupling between the
cavity and reservoir (see Eq.\ref{EqHex}), $\wp(\omega)$ is the
density of states in the reservoir,
$a_{if}=\langle\psi_i|a|\psi_f\rangle$. The Lorentzian factor
accounts for the broadening of the state, $\gamma_{ph}$, provided
by finite lifetime of cavity photons.

If the system is in the mixed state, the spectrum (\ref{Iomega1})
can be estimated as a sum over all possible initial states taken
with corresponding probabilities $P_i$,
\begin{equation}
S(\omega)\sim\frac{2\pi}{\hbar}\sum_{if}P_i|a_{fi}|^2\wp(\omega)\frac{\gamma_{ph}^2}{(\varepsilon_f-\varepsilon_i-\hbar\omega)^2+\gamma_{ph}^2}.
\label{EqFermiGoldenRule}
\end{equation}
Parameters $P_i$ entering into the above expression are nothing
but diagonal matrix elements of the density matrix in the
stationary state written in the basis of eigenstates of the
Hamiltonian (\ref{EqFullHamiltonian}),
$P_i=\tilde{\rho}_{ii}^{SS}$. The matrix $\tilde{\rho}^ {SS}$ can
be found by an unitary transformation of the stationary density
matrix in the basis $(|e,n\rangle,|g,n\rangle)$ obtained from
solution of Eqs.~(\ref{EqLindbladMatrix}.)

The described phenomenological approach for calculating the
emission spectrum has one substantial drawback. Namely, we assume
the peaks in the spectrum to be Lorentzians which is not always
guaranteed \cite{Scully_b01,LibLaussyPRB2011,LibDelVallePRA2011}.
In general, according  to the Wiener-Khintchine theorem
\cite{LibCarmichael}, the spectrum of the emission from the system
can be calculated as a Fourier transform of two-time correlator,
\begin{equation}
S(\omega)\sim\lim_{t\rightarrow\infty}{\cal{R}}\int_0^\infty\langle
a^\dagger(t+\tau)a(\tau)\rangle e^{i\omega\tau}d\tau.
\end{equation}
Generally, the calculation of two-time correlators is a
complicated task which cannot be solved exactly. However, it is
well-known from literature \cite{Scully_b01,LibCarmichael} that
the using of certain general assumptions about the behavior of the
system allows to reduce the calculation of two-time correlators to
calculation of one-time correlators within the framework of the
the quantum regression theorem (QRT) \cite{QRTbib}. This means
that the spectrum can be calculated straightforwardly from the
density matrix of the system in stationary state. For the system
we consider in the present paper, the approach based in QRT
results in
\begin{equation}
S(\omega)=\frac{1}{\pi}{\cal
R}\sum_{i,j,k,l;a,b}[({M}+i\hbar\omega
{I})^{-1}]_{ij,kl}^{a,b}(\rho_{km}^{ab(SS)})a_{lm}a_{ji},
\label{EqSpectrumQRT}
\end{equation}
where $ M$ is the matrix defined in Eqs.~(\ref{EqLindbladMatrix}),
(\ref{EqLindbladMatrixExplicit}), $I$ is the unity matrix. Iy
should be noted that the expression above is valid for any choice
of the basis.

The shape of the emission spectrum of the system calculated using
the Fermi golden rule and QRT is analyzed in the following
section.

\section{Discussion}

\begin{figure}
\includegraphics[width=1.0\linewidth]{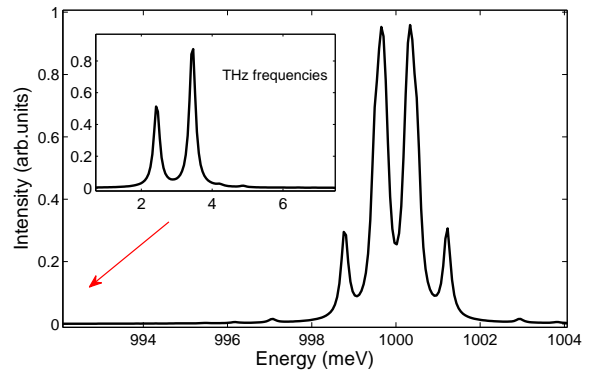}
\caption{Emission spectrum of the asymmetric QD- cavity system
calculated using the Fermi golden rule. In optical diapason close
to the eigenfrequency of the cavity standard quadruplet structure
is revealed. In the range of frequencies close to the Rabi
frequency additional set of emission peaks appears. They are
provided by asymmetry of the QD and are absent in a symmetric
case. The parameters of the calculation are:
$\gamma_{ph}=\gamma_{QD}=0.1$ meV; $P\approx 0.15$ meV;
$\hbar\omega_c=\Delta=1$ eV; $g_R\approx g_S\approx 1$ meV. }
\label{FigArbFermi}
\end{figure}

The emission spectrum of the system, calculated using the
phenomenological approach based on the Fermi golden rule, is
presented in Fig.~\ref{FigArbFermi}.

\begin{figure}
\includegraphics[width=1.0\linewidth]{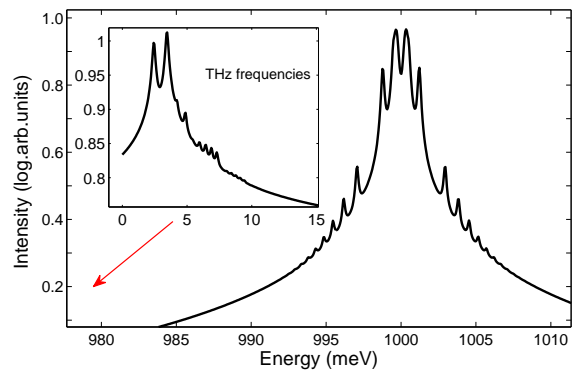}
\caption{Emission spectrum of the asymmetric QD- cavity system
calculated using Fermi golden rule in logarithmic scale. Multiplet
structure of the emission around eigenfrequency of the cavity
becomes more visible then in the linear scale. In the range of the
frequencies close to Rabi frequency additional set of emission
peaks appears. Parameters of the system are the same as for
Fig.~\ref{FigArbFermi}} \label{FigLogFermi}
\end{figure}

One sees that in the region of the frequencies close to the eigen
frequency of the cavity (coinciding with the frequency of optical
transition in QD, $\hbar\omega_c=\Delta$) the spectrum reveals a
quadruplet pattern. The appearance of the additional multiplets is
not visible in the linear scale, but becomes apparent in
logarithmic scale as it is seen in Fig.~\ref{FigLogFermi}. These
results are in good qualitative agreement with those obtained
earlier for the symmetric QDs in strong coupling regime
\cite{LibLaussyFermions}. This is not surprising, since main
difference in the emission from symmetric and asymmetric QDs
appears at the regions around the Rabi frequency and the double
frequency $2\omega_c$.

The insets in Figs.~\ref{FigArbFermi},\ref{FigLogFermi} show the
emission pattern at THz range about $\Omega_R$. In full agreement
with results of the Section II, one sees the appearance of the
peaks in the emission at frequencies $\omega=\Omega_R\sqrt{n}$
with $n=1,2,3,...$. It should be noted that one should expect very
low intensities of THz emission as compare to the emission in the
optical diapason, since density of states of the photon reservoir
scales  as $\omega^3$. However, the situation can be improved if
the coupled QD-cavity system is placed inside a bigger cavity
tuned at the THz range. In this case the density of states has a
sharp peak around the eigenfrequency of THz cavity and the rate of
spontaneous emission is dramatically increased due to the Purcell
effect \cite{LibPurcell}. Current state of technology allows the
increasing of the emission rates in the THz regime by a factor of
$100$ in high-quality cavities \cite{Todorov}.

\begin{figure}
\includegraphics[width=1.0\linewidth]{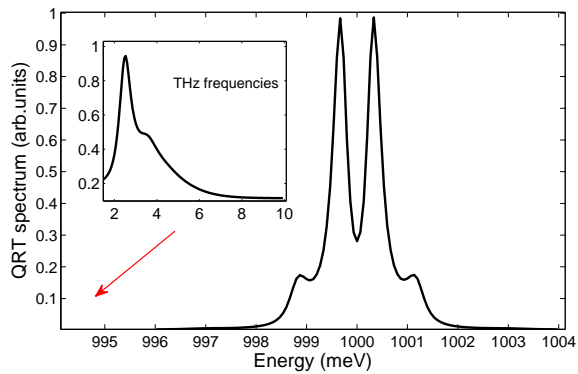}
\caption{Emission spectrum of the asymmetric QD- cavity system
calculated using QRT. In optical diapason close to the
eigenfrequancy of the cavity standard quadruplet structure is
revealed. In the range of the frequencies close to Rabi frequency
a single broad asymmetric emission peak appears. The parameters of
the calculation are the same as in Fig.~\ref{FigArbFermi}}
\label{FigArbQRT}
\end{figure}

The results of the calculation of the spectrum based on using the
quantum regression theorem are presented in Fig.~\ref{FigArbQRT}.
The part of the spectrum corresponding to the optical diapason for
the frequencies about $\omega\approx\omega_c$ is in good agreement
with results obtained by using the Fermi golden rule
(Figs.~\ref{FigArbFermi}, \ref{FigLogFermi}). The differences are
that the side peaks in the QRT plots have lower intensities, and
the shape of the spectrum is more smooth, so that multiplet
pattern becomes invisible.

The changes in the THz part are more dramatic. Instead of series
of the narrow peaks shown at the inset of Fig.~\ref{FigArbQRT} one
sees a formation of a single broad strongly assymmetric peak
centered at a frequency $\omega\approx\Omega_R$. However,
qualitative result remains the same: new optical transitions at
THz range are opened by the asymmetry of the QD.

The important question is statistical properties of the emitted
THz light. As emission spectrum we obtain is quite large, and
efficiency of THz emission is normally suppressed as compare to
the emission at optical frequencies, one can expect that emitted
radiation will have thermal statistics with second order coherence
$g^ {(2)}\approx 2$. On the other hand, placing of the sample into
high quality THz cavity can lead to selection of more narrow
region of the frequencies of the emission. In this case one can
expect the possibility to achieve THz lasing regime with $g^
{(2)}\approx 1$. The detailed consideration of this situation,
however, lies beyond the scopes of the present paper. As to
detection of the emitted THz radiation, it can be achieved by
standard THz detectors (see, e.g., the review \cite{Sizov10}).

\section{Conclusions}

In conclusion, we considered an asymmetric two-level quantum
system corresponding to small asymmetric QD interacting with an
electric field of a single-mode microcavity. We found analytical
solutions for the eigenenergies of the system for the cases of
weak asymmetry and high photon occupation numbers. As well, we
developed numerical approach for the calculation of the emission
spectrum under incoherent pumping. It is shown that in the regime
of strong pump a new set of peaks in the emission appears in the
regions close to the Rabi frequency and double transition
frequency. This allows to use the proposed system as nonlinear
optical element and tunable source of the THz radiation.

The work was partially supported by Rannis "Center of excellence
in polaritonics", the RFBR Project No. 10-02-00077, the Russian
Ministry of Education and Science, the 7th European Framework
Programme (Grants Nos. FP7-230778, FP7-246912 and FP7-246784), and
ISTC Project No. B-1708. We thank Dr. F. Laussy for useful
discussions and International Institute of Physics (Natal, Brazil)
for hospitality.

\end{document}